 \documentclass[
  pre,
  aps,
  a4paper,
  english,
  reprint,
  twocolumn,
  showpacs,
  showkeys,
  superscriptaddress
 ]{revtex4-1}
\usepackage{tikz}
\usepackage{amsmath}
\usepackage{blkarray}
\usepackage{graphicx,subfigure}
\usepackage{amssymb}
\usepackage{amsmath}
\usepackage{mathrsfs}
\usepackage{multirow}
\usepackage{nicematrix}
\usepackage{pifont}
\usepackage{epstopdf}
\usepackage{hyperref}
\usepackage[mathscr]{euscript}
\usepackage[english]{babel}
\usepackage{setspace}

\usepackage[utf8]{inputenc}
\usepackage[T1]{fontenc}

\usepackage{qcircuit}

\begin{document}

\title{Repeated quantum game as a stochastic game: Effects of the shadow of the future and entanglement}

\author{Archan Mukhopadhyay}
\email{archan91@gmail.com \& a.mukhopadhyay@bham.ac.uk}
\affiliation{Centre for Brain and Mind, Department of Psychiatry, National Institute of Mental Health and Neurosciences, Bangalore 560029, India}
\affiliation{School of Computer Science, University of Birmingham, Birmingham B152TT, United Kingdom}
\author{Saikat Sur}
\email{saikat.sur@weizmann.ac.il (corresponding author)}
\affiliation{Department of Chemical and Biological Physics \& Department of AMOS, Weizmann Institute of Science, Rehovot 7610001, Israel}
\author{Tanay Saha}
\email{tanays@iitk.ac.in}
\affiliation{Department of Physics, Indian Institute of Technology Kanpur, Uttar Pradesh 208016, India}
\author{Shubhadeep Sadhukhan}
\email{shubhadeep.sadhukhan@weizmann.ac.il}
\affiliation{Department of
Chemical and Biological Physics, Weizmann Institute of Science, Rehovot 7610001, Israel}
\author{Sagar Chakraborty}
\email{sagarc@iitk.ac.in}
\affiliation{Department of Physics, Indian Institute of Technology Kanpur, Uttar Pradesh 208016, India}
\date{\today}
\begin{abstract}
We present a systematic investigation of the quantum games, constructed using a novel repeated game protocol, when played repeatedly ad infinitum. We focus on establishing that such repeated games---by virtue of inherent quantum-mechanical randomness---can be mapped to the paradigm of stochastic games. Subsequently, using the setup of two-player--two-action games, we explore the pure reactive strategies belonging to the set of reactive strategies, whose support in the quantum games is no longer countably finite but rather non-denumerably infinite. We find that how two pure strategies fare against each other is crucially dependent on the discount factor (the probability of occurrence of every subsequent round) and how much entangled the quantum states of the players are. We contrast the results obtained with the corresponding results in the classical setup and find fundamental differences between them: e.g, when the underlying game is the prisoner's dilemma, in the quantum game setup, always-defect strategy can be beaten by the tit-for-tat strategy for  high enough discount factor.
\end{abstract}

\maketitle


\section{Introduction}
Players, their actions and strategies, and the payoffs they get constitute basic ingredients of any game~\cite{Neumann_Morgenstern_2007}. The mechanism of realizing the payoffs by the players from the point of adoption of the strategies is usually not of much interest to classical game theorists: They find the equilibrium solutions for a given payoff structure. However, if the aforementioned mechanism is quantum-mechanical~\cite{eisert1999prl, eisert1999jmodopt, pitrowski2003, pitrowski2002physica, pitrowski2003physica}, then it changes a game fundamentally: new actions and strategies are facilitated that are otherwise inaccessible. In fact, in the paradigmatic game of Prisoner's Dilemma (PD), it was shown that efficient pareto-optimal Nash equilibrium (NE) strategy~\cite{eisert1999prl, eisert1999jmodopt}---known as the super-cooperation strategy---exists when one deals with a quantum-mechanical version of it.

It may be recalled, in this context, that PD has long puzzled game theorists interested in the causes behind the evolution of cooperation. The dilemma is in the fact that although the reward of mutual cooperation is far more, the mutual defection is arguably the rational choice as any unilateral cooperation by a player fetches her lower payoff~\cite{Luce_Raiffa_1957}. Thus, even in this extremely simple game setup, the possible emergence of cooperation is an enigma. It however was established through experiments~\cite{Axelrod_1980} that the strategies adopted while playing the game repeatedly may establish cooperation. To illustrate the strategies in repeated games, one can think about the example of a strategy that always cooperates (say, ALLC) is dominated by a strategy that always chooses to defect (ALLD). There is also another celebrated strategy, Tit-for-Tat (TFT), that cooperates only in the first round and then mimics its opponent from the next round onwards. This implies that TFT against TFT always results in mutual cooperation. But always-defect strategy (ALLD) exploits TFT only in the first round, then both receive the payoff corresponding to the mutual defection thereon. 

\par Till now the study of repeated PD games has mainly evolved around classical games. However, with the advent of quantum information technologies, we are slowly approaching a society where the presence of quantum systems will be 
rich~\cite{eisert1999prl,eisert1999jmodopt, khan2018qip,benjamin2001pra,enk2002pra,flitney2006pla,szopa2014ekon,li2014sr,iqbal2016prsb,luna2017sr,phoenix2020pla}. Thus, the quantum game, on its own merit, is gaining importance with time. The definitions and motivations of quantum games are not unique: It is still in development phase~\cite{Li2013SciRep,Li2013_PLosOne}. The literature has majorly focused on one-shot quantum games. The literature on repeated quantum games is still in its nascent phase. There have been very few studies in this segment: Iqbal and Toor introduced a repeated quantum game protocol for the first time in the literature~\cite{iqbal_toor}; and Frckiewicz revised this protocol and showed that a quantum repeated game can give better results compared to the classical counterpart~\cite{Frckiewicz2021}. Ikeda and Aoki~\cite{Ikeda2020} discussed the equilibrium concept in a particular repeated quantum game setup where the quantum strategies are restricted to a space spanned by the Pauli strategies. However, to best of our knowledge, a study regarding the effectiveness of known equilibrium strategies of a repeated game like TFT, ALLD, and  ALLC~\cite{axelrod1984book, nowak2003book, hilbe2013pnas} in the quantum domain is missing in literature. 

The setup of the repeated quantum game is not unique. It is always possible to construct a more apt setup, like the Eisert--Wilkens--Lewenstein (EWL) protocol~\cite{eisert1999prl}, that is closer to the known protocols of classical repeated games. Obviously, the landscape of strategies in the repeated PD is much richer in quantum games. Moreover, since super-cooperation is even more efficient than cooperation, how it fares against defection in repeated game set-up is of as much interest as any cooperation-inducing strategies. The results are furthermore expected to depend on the shadow of the future~\cite{BS}, i.e., the probability of occurrence of the subsequent round. Even more interestingly, quantum entanglement can play a very important role---an aspect completely absent in the classical games. In this paper, we aim to present a systematic presentation and investigation of the basic paradigm of quantum repeated games, and contrast the results obtained with those available in the corresponding classical repeated games.

One of the main achievements of this work is crystal clearly bringing forth the fact that a repeated quantum game is inherently a stochastic game. Stochastic game~\cite{Solan_Vieille_2015}, first introduced in 1953~\cite{Shapley_1953}, dynamically takes into account the possible changes in game environment due to action of the players. Specifically, the payoff matrix that is supposed to incorporate environmental effects, changes in the subsequent round of play in response to the decisions taken by the players in the present round. This is unlike the classical repeated game alluded to in preceding paragraphs: Usually, in such repeated games---studied within the paradigm of evolutionary game theory~\cite{fundenberg1990, bo2011}---the underlying game remains constant. Of course, one can explicitly model in the changing payoff matrix (and hence, effectively the underlying game) into such setup~\cite{Hilbe2018} and find interesting results, e.g., enforcing cooperation via such stochasticity. However, as elaborated later in the paper, in repeated quantum games, there is no need to model in stochasticity separately: It is built-in through the EWL protocol. This gives rise to fundamentally different results, e.g., unlike the classical case~\cite{nowak_evolutionary_dynamics}, ALLD is no longer always unbeatable by TFT.

\section{Repeated Quantum Game: Setup}
We now set up the mathematical framework, that is backbone of our paper, while succinctly presenting some of the indispensable known concepts for the sake of clarity and completeness.
\subsection{Classical Game}
\label{sec:CG}
The standard normal form representation of a two-player game, equipped with two classical strategies, say, Cooperate ($C$) and Defect ($D$), is given as
\begin{eqnarray*}  
	\centering
    \begin{tabular}{cc|c|c|}
      & \multicolumn{1}{c}{} & \multicolumn{2}{c}{Bob}\\
      & \multicolumn{1}{c}{} & \multicolumn{1}{c}{$C$}  & \multicolumn{1}{c}{$D$} \\\cline{3-4}
      \multirow{2}*{Alice}  & $C$ & $R,R$ & $S,T$ \\\cline{3-4}
      & $D$ & $T,S$ & $P,P$ \\\cline{3-4}
    \end{tabular}
    \label{table:payoff}
\end{eqnarray*}
where the first element of each entry is the payoff of Alice, a player, and the second element is that of another player, Bob, say. The payoff elements $R$, $S$, $T$, and $P$ are popularly known as Reward, Sucker's payoff, Temptation, and Punishment respectively. As the game is symmetric, from now on, we shall only focus on the payoff matrix, to be denoted by ${\sf A}$, of Alice. For the game matrix to correspond to the PD, it should be imposed that $T>R>P>S$ and $2R>S+T$.

\begin{figure}
 \includegraphics[width=0.25\textwidth]{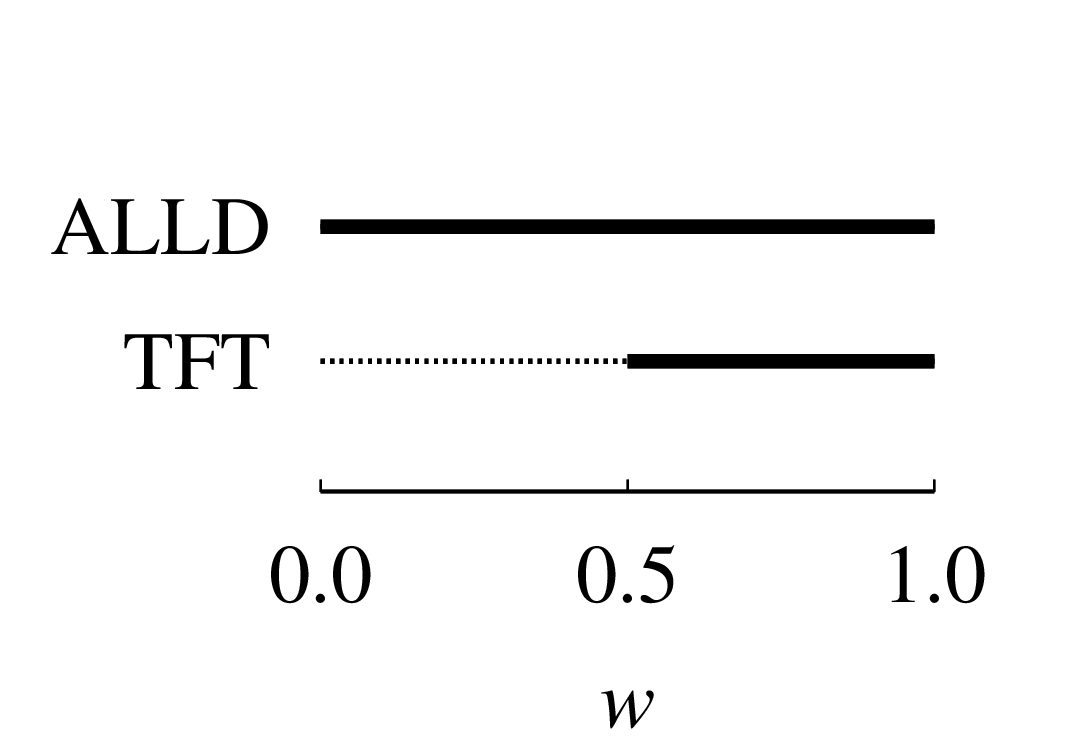}
 \caption{ALLD is always unbeatable: {Schematic presentation of the range (solid black line) of discount factor $w$ for which TFT and ALLD are  strict NE's in the classical repeated PD game. For illustrative purpose, we have have fixed $R=3, S=0, T=5,\,{\rm and}\, P=1$.}}
 \label{fig_classical} 
 \end{figure}

Suppose this game is played repeated ad infinitum, such that at every round of play, a player reacts to its opponent's immediately preceding action, then strategies like ALLC, ALLD, and TFT can be played. If the probability of occurrence of the subsequent round is $w$ (also called the discount factor), then one arrives at~\cite{nowak_evolutionary_dynamics}
\begin{eqnarray*}  
	\centering
    \begin{tabular}{cc|c|c|}
      & \multicolumn{1}{c}{} & \multicolumn{2}{c}{}\\
      & \multicolumn{1}{c}{} & \multicolumn{1}{c}{TFT}  & \multicolumn{1}{c}{ALLD} \\\cline{3-4}
    \multirow{2}*{  {${\sf A}$}=}  & TFT & $R/(1-w)$ & $S+wP/(1-w)$ \\\cline{3-4}
      & ALLD & $T+wP/(1-w)$ & $P/(1-w)$ \\\cline{3-4}
    \end{tabular}
    \label{table:TFT-ALLD}
\end{eqnarray*}
for strategic interaction between two players through TFT and ALLD. The rational outcome is theorized to be the Nash equilibrium (NE)---the strategy profile that ensures that unilateral deviation from it results in no gain~\cite{nash1950pnas} for the deviating player. Thus, TFT is a strict NE of the game if ${\sf A}_{\textnormal{TFT},\textnormal{TFT}}> {\sf A}_{\textnormal{ALLD},\textnormal{TFT}}$ and likewise for ALLD to be strict NE, it is required that ${\sf A}_{\textnormal{ALLD},\textnormal{ALLD}}> {\sf A}_{\textnormal{TFT},\textnormal{ALLD}}$. Here ${\sf A}_{i,j}$ (also written as ${\sf A}_{ij}$ when no ambiguity arises) is the $ij$-th element of ${\sf A}$. It is quite straightforward to see that TFT is a strict NE given $w>(T-R)/(T-P)$, while ALLD is always strict NE. We depict this known result schematically in  Fig.~\ref{fig_classical}.
 \subsection{Quantum Game}
 \label{sec:QG}
 The extension of the one-shot game into the quantum domain maybe done using the EWL protocol~\cite{eisert1999prl} that maps the classical strategies into a continuum set of quantum strategies available to the players. The main idea behind the extension is to arrive at  the exact outcomes of the classical strategies $C$ and $D$ through manipulations of Hilbert state vectors $|{0}\rangle$ and $|{1}\rangle$ respectively. One defines state of the quantum game as $|\psi\rangle\equiv{\sf J}(\varepsilon)|ab\rangle={\sf J}(\varepsilon)|a\rangle_A\otimes|b\rangle_B$ where $a,b\in\{0,1\}$ and the unitary entangling operator
 \begin{equation}
\sf{J}(\varepsilon)  \equiv {\sf 1}_A\otimes{\sf 1}_B \cos\varepsilon  +\sqrt{-1}\, {\sf X}_A \otimes {\sf X}_B \sin\varepsilon.
 \label{eq:J}
\end{equation}
 ($A$ and $B$ correspond to Alice and Bob respectively.) Here, $\varepsilon$ is the parameter characterizing the entanglement between the qubits, $\sf{1}$ is the identity operator and ${\sf X}$ is the standard qubit flip operator or quantum NOT gate. The action of ${\sf X}$ on the state $|0\rangle$  takes it to its orthogonal counterpart $|1\rangle$ and vice versa. The two-qubit entangling operator ${\sf J}(\varepsilon)$, which belongs to the joint Hilbert space $\mathcal{H}_A \otimes \mathcal{H}_B$, forms an  entangled two-qubit state out of the pure states $|0\rangle_A$ and $|0\rangle_B$. The entanglement parameter, $\varepsilon \in [0,\pi/2]$, determines the entanglement of the initial state given to the players. The case $\varepsilon = \pi/4$ corresponds to the maximum entanglement between the states of the players, while $\varepsilon = \pi/2$  and $\varepsilon = 0$ correspond to zero entanglement. 

Any strategy now is recognized as a unitary operator belonging to a $SU(2)$ group rotation is of the form
\begin{eqnarray}
 {\sf U}=
\begin{pmatrix} 
 \gamma & \delta \\
-\delta^* & \gamma^*
\end{pmatrix}
\label{eq:3}
\end{eqnarray}
with $|\gamma|^2 + |\delta|^2 = 1$; the elements are complex numbers and asterisk denotes the complex conjugation operation. We shall use the superscript $A$ or $B$ to specify whether Alice or Bob has played the strategy. It is customary to start with initial game state as $|\psi_{\rm initial}\rangle\equiv\sf{J}(\varepsilon)|{00}\rangle$. The simultaneous play by Alice and Bob is constructed to lead to a new final state $|\psi_{\rm final}\rangle = {\sf J}^\dagger (\varepsilon) ({\sf U}^A \otimes {\sf U}^B) {\sf J} (\varepsilon) |00\rangle$. The payoff is defined to be given by $\$_A\equiv \sum_a\sum_b{\sf A}_{ab}|\langle ab|\psi_{\rm final}\rangle|^2$ which justifies the classical to quantum extension. This is because with ${\sf U}$ as ${\sf 1}$ and ${\sf X}$---respectively recognized as cooperation and defection actions---the payoff is identical to what the classical game with the two actions would yield.

Finally, we introduce a protocol, a generalization of the EWL protocol, which can be used to play a repeated  quantum game. Before going into the details of this protocol, we would like to emphasize a possibly obvious fact that the players have the knowledge of the actions and payoffs but are ignorant about the quantum states and the detailed quantum operations occurring in the game. Here the repeated version of the PD game is initialized with $|00\rangle$ basis state (same as in the EWL protocol). Since ${\sf J}^\dagger {\sf J}={\sf 1} $, the final state obtained after $m$ iterations of one-shot games is as follows:

\begin{eqnarray}
|\psi_m\rangle = {\sf J}^\dagger (\varepsilon) \prod^m_{k=1}({\sf U}^A_k \otimes {\sf U}^B_k) {\sf J} (\varepsilon) |00\rangle,
\label{eq: psi_n}
\end{eqnarray} %
 where, ${\sf U}^A_k$ and ${\sf U}^B_k$ are the quantum strategies played by the players $A$ and $B$ at the $k$ th iteration/round of the game. The cumulative payoff for Alice is the sum (weighted by the discount factor) of the payoffs obtained after each iteration, i.e.,
 \begin{eqnarray}
\$_{A} \equiv \lim_{M\to\infty}\sum^{M}_{m=1} \sum_{b=0}^1 \sum_{a=0}^1w^{m-1} {\sf A}_{ab} |\langle ab | \psi_m \rangle|^2, 
 \label{eq: cumulative_payoff}
 \end{eqnarray}
  with $w\in [0, 1)$; Bob's payoff follows similarly. The existence of the above limit can be trivially shown by employing the comparison test using series $\sum_{m=1}^{M \to \infty}w^{m-1}T \equiv T/(1-w)$: Obviously, $0 \le \sum_{b=0}^1 \sum_{a=0}^1w^{m-1} {\sf A}_{ab} |\langle ab | \psi_m \rangle|^2 \le w^{m-1}T~\forall m$, hence the limit in Eq.~(\ref{eq: cumulative_payoff}) converges. 
  
  A close inspection of Eq.~\ref{eq: psi_n} reveals that the entanglement dependency of the projections of the final state on the basis states comes through terms like $\sin{\varepsilon}\cos{\varepsilon}$, ${\sin^2{\varepsilon}}$, and/or  ${\cos^2{\varepsilon}}$. Therefore, the final state at $\varepsilon=\theta$ is identical with that at $\varepsilon=\pi/2-\theta$. It connotes that it is sufficient to explore the cases with $\varepsilon$ within the range $[0,\pi/4]$ where entanglement is monotonically increasing.
  \subsection{Hint of Stochastic Game}

       \begin{figure*}
   \begin{center}
   \includegraphics[width=0.98\textwidth]{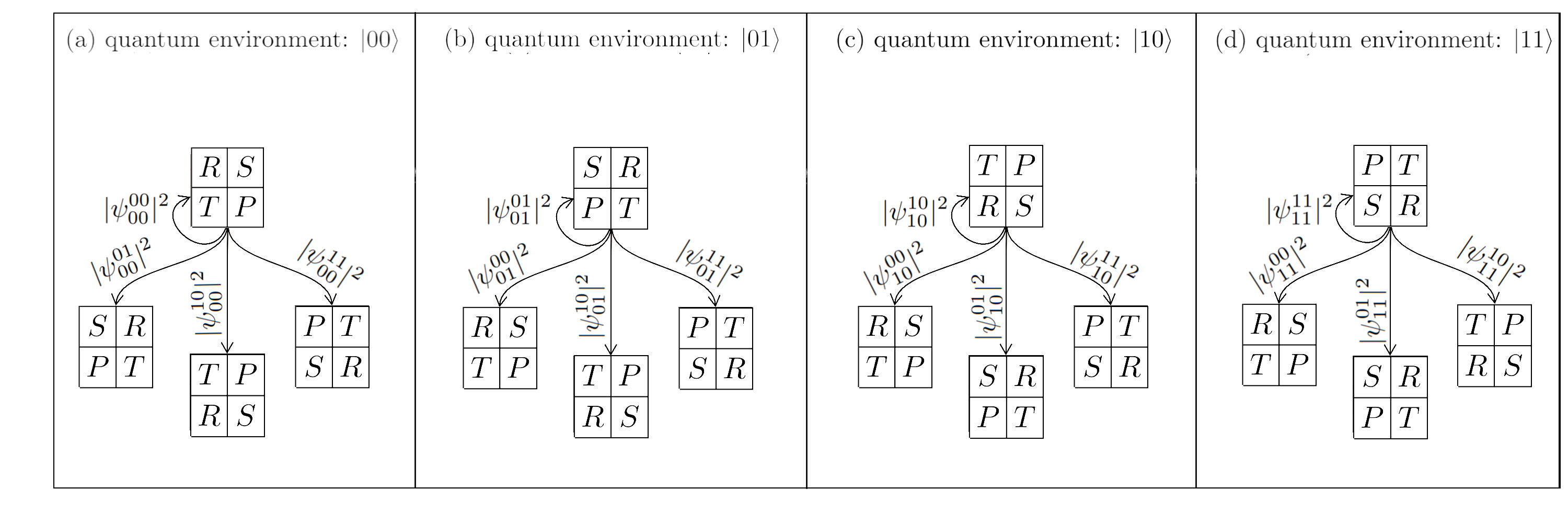}
    \end{center}
   \caption{{Repeated quantum game as stochastic game: Schematic presentation of the stochastic transitions to different quantum environments (equivalently represented by the payoff matrices) from an initial quantum environment with whom the central payoff matrix in each subfigure corresponds to. Here the transition probabilities are dependent on the propagators: $\psi^{a'b'}_{ab} \equiv\langle a'b'\vert {\sf J}^\dagger (\varepsilon) ({\sf U}^A \otimes {\sf U}^B) {\sf J} (\varepsilon) |ab\rangle$. }}%
  \label{fig:sg} 
   \end{figure*} 
The quantization scheme of a classical game is, by construction, initial-state dependent. A careful introspection about the quantum strategy protocol reveals to us that the definition of conventional strategies like ${\sf C}$ or ${\sf D}$, through the corresponding unitary operators, is motivated by the fact that the initial state (before entanglement) is $\vert 00 \rangle$. This sets up the environment of the game being player; in fact, let's agree to call the initial state---which is entangled by $\sf J$ and subsequently operated by the unitary operators corresponding to the actions available to the players---`quantum environment': This terminology's appositeness will become clearer as we discuss further in this section. 

Observe that had the initial state (before entanglement) been $\vert 11 \rangle$, say, then the definitions of ${\sf C}$ and ${\sf D}$ would have to be interchanged so as to keep the payoff matrix intact as in classical two-player--two-action PD game with cooperation and defection actions. From a different angle, if we choose not to change the definitions of ${\sf C}$ or ${\sf D}$ then the payoff matrix of classical two-player--two-action PD game would come out to be something different via EWL protocol. This is illustrated in Table~\ref{tab:stochastic_payoff}.
\begin{table}[ht]
\centering
\begin{tabular}{ |p{2cm}|p{2cm}|p{2cm}|p{2cm}|  }
 \hline\hline
~~~~~~$\vert 00 \rangle$ & ~~~~~~$\vert 01 \rangle$ & ~~~~~~$\vert 10 \rangle$ & ~~~~~~$\vert 11 \rangle$ \\
 \hline\hline
{$\left[\begin{array}{cc}
R & S  	\\
T & P 	
\end{array}\right]$}  & {$\left[\begin{array}{cc}
S & R  	\\
P & T 	
\end{array}\right]$} & {$\left[\begin{array}{cc}
T & P  	\\
R & S 	
\end{array}\right]$} &  {$\left[\begin{array}{cc}
P & T  	\\
S & R 	
\end{array}\right]$}\\
  \hline\hline
\end{tabular}
\caption{State dependent payoffs hinting towards a resemblance between quantum game and stochastic games: This table illustrates that if the initial state is $ab$ where $a,b\in\{0,1\}$, and the cooperation and the defection strategies correspond to $\sf C$ and $\sf D$ operators, then what the corresponding classical payoff matrix for Alice would be for two-player--two-action one-shot game with cooperation and defection as possible actions.}
\label{tab:stochastic_payoff}
\end{table}
With this in mind, we recall that in the repeated quantum game, every player is unaware of the present quantum environment except the initial one; the only things she knows is her quantum strategy (fixed at the initial round), the initial state, and the cumulative payoff she earns after the game is complete. Since the quantum environment available to the player after each round is the one (out of $\vert 00 \rangle$, $\vert 01 \rangle$, $\vert 10 \rangle$, or $\vert 11 \rangle$) onto which the final quantum state has collapsed post measurement needed for payoff calculation for the round, the quantum environment is not necessarily specified by the initial $\vert 00 \rangle$ state but any one of the four basis states.

In other words, we can say that the repeated quantum game is nothing but a stochastic game by virtue of its inherent quantum-mechanical randomness. Like the stochastic games, it models the dynamic interactions in which the (quantum) environment evolves in response to players' actions. The game matrix changes at every round: One of the four payoff matrices has to be considered with a probability that depends in the quantum protocol. We schematically present this transition between the game matrices in Fig.~\ref{fig:sg}. We remark that this interesting phenomenon forms the backbone for the non-trivial results that this paper showcases.

\subsection{The Strategies}
\label{sec: strategies}
While in one-shot games, action and strategy are used synonymously, in the case of repeated games the distinction between them must be made. A strategy in the repeated game is the entire sequence (e.g., TFT, ALLD , or ALLC) of actions ($C$ or $D$ in classical setting) played by a player. In the quantum games, the set of actions become uncountably infinite as explained below.

As mentioned earlier, after the entangled state is prepared, the players are allowed to choose their actions from their respective action set---and perform local operations on the joint entangled state.  Let the operators ${\sf U}_A$ and ${\sf U}_B$ represent the local single-qubit unitary gate operations performed by the players, Alice and Bob, respectively. Finally, a two qubit disentangling operator $\sf{J}^\dagger (\varepsilon) = {\sf 1}_A\otimes{\sf 1}_B \cos\varepsilon  -i {\sf X}_A \otimes {\sf X}_B \sin\varepsilon$ acting on the entangled state produces a pure state $\vert \psi \rangle$ with individual states $|\psi\rangle_A$ and $|\psi\rangle_B$ for the players Alice and Bob respectively. Since both the input and the output states of a unitary operation lie on the surface of the Bloch sphere, the  general form of a unitary operation belonging to a $SU(2)$ group rotation is of the form given by Eq.~(\ref{eq:3}). The most general form~\cite{nielsen2000book} of the local unitary operation---belonging to $SU(2)$ group---is given, most generally, in terms of three  parameters $\theta$, $\alpha$ and $\phi$ such that $\gamma = \cos\frac{\theta}{2} - i \cos \alpha \sin\frac{\theta}{2}$ and $\delta = -i e^{-i \phi} \sin \alpha \sin{\frac{\theta}{2}}$. The unitary operator rotates the input state by an angle $\theta$ about the $(\sin \alpha \cos \phi, \sin \alpha \sin \phi,  \cos \alpha)$ axis of the Bloch sphere. Thus, the quantum action space can be said to be spanned by the parameters $\theta$, $\alpha$ and $\phi$ where $\theta \in [0,2\pi)$, $\alpha \in [0,\pi)$ and $\phi \in [0,2\pi)$. In summary, we can specify the quantum actions for Alice and Bob as a pair of triplets ${\bf \Theta}^A=\{\theta_A, \alpha_A, \phi_A\}$ and ${{\bf \Theta}^B}=\{\theta_B, \alpha_B, \phi_B\}$ respectively.  The action counterpart of classical cooperation and defection is given by the unitary operators with $(\theta = 0)$ and $(\theta = \pi, \alpha = \pi/2, \phi = 0)$.  

In this paper, we are interested in mostly finding how tit-for-tat strategy, all-quantum strategy, and all-classical strategy fares against each other (see Appendices~\ref{appendix:B}~and~\ref{appendix:C} for detailed analysis). For the sake of concreteness and clarity of the results, we have restricted our main text to the study of four interesting actions which are tabulated in Table~\ref{tab:strategies_3}. Apart from the quantum counterpart of cooperation and defection, here we have super-cooperation~~\cite{eisert1999prl} action ${\sf Q}$ that is a Pareto-optimal NE in the one-shot two-player three-strategy game; it fetches the same payoff as ${\sf C}$. Furthermore, we pick another well-known action, namely, the dephased Hadamard gate~\cite{nielsen2000book}, ${\sf H}$, to illustrate our results.
\begin{table}[ht]
\centering
\begin{tabular}{ |p{3.75cm}||p{1cm}|p{1.5cm}|p{1.5cm}|  }
 \hline\hline
Actions & $\theta$ & $\alpha$ & $\phi$ \\
 \hline\hline
{Cooperation, ${\sf C}$}  & $0$ & ${\rm arbitrary}$ &  ${\rm arbitrary}$\\
 {Defection, ${\sf D}$ }&   $\pi$  & $\pi/2$   & $0$\\
{Super-cooperation, ${\sf Q}$  }   & $\pi$    & $0$&   ${\rm arbitrary}$\\
 {Dephased Hadamard, ${\sf H}$  }  & $\pi$ & $\pi/4$&  $\pi/2$\\
  \hline\hline
\end{tabular}
\caption{Specification of some standard actions in the three-parameter space.}
\label{tab:strategies_3}
\end{table}
The payoff matrix of a one-shot interaction among these four strategies has the following payoff matrix:
\begin{widetext}
\centering
    \newcommand{\myline}{\cline{2-5}}
    \newcommand{\noline}[1]{\multicolumn{1}{c}{#1}}
    \begin{tabular}{c|c|c|c|c|}
        \noline{} & \noline{ $ {\sf C} $ } & \noline{$ {\sf D} $} & \noline{$ {\sf Q} $} & \noline{$ {\sf H} $} \\
        \myline
        $ {\sf C} $ & $R$ & $S$ & $R \cos^2 2\varepsilon + P \sin^2 2\varepsilon$ & $\frac{(R+S)\cos^2 2\varepsilon + (T+P)\sin^2 2\varepsilon}{2}$ \\
        \myline
        $ {\sf D} $ & $T$ & $P$ & $T \cos^2 2\varepsilon + S \sin^2 2\varepsilon$ & $\frac{(R+S)\sin^2 2\varepsilon + (T+P)\cos^2 2\varepsilon}{2}$\\
        \myline
        $ {\sf Q} $ &  $R \cos^2 2\varepsilon + P \sin^2 2\varepsilon$ & $S \cos^2 2\varepsilon + T \sin^2 2\varepsilon$ & $R$ & $\frac{R+S}{2}$\\
        \myline
        $ {\sf H} $ & $\frac{(R+T)\cos^2 2\varepsilon + (S+P)\sin^2 2\varepsilon}{2}$ & $\frac{(R+T)\sin^2 2\varepsilon + (S+P)\cos^2 2\varepsilon}{2}$ & $\frac{R+T}{2}$ & $\frac{R+S+T+P}{4}$ \\
        \myline
    \end{tabular}
    \centering
\end{widetext}
where, as an aside, it is interesting to note that in the presence of $\sf H$, i.e., in the two-player-four-action game, super-cooperation is no longer NE.

Coming on the strategies, we are thus naturally interested in ALLC and ALLD strategies as has been the case with classical games. Moreover, now ALLQ (always super-cooperation, ${\sf Q}$) and ALLH (${\sf H}$) are of immediate interest as well. As is the case with TFT, in the context of classical games, here we can analogously construct two tit-for-tat strategies that can help us understand evolution and efficiency of supercooperation and cooperation. This is because a player adopting TFT has two options while starting the repeated game: either to start the first round with cooperation (${\sf C}$) or to start with supercooperation (${\sf Q}$). We term the first one as CTFT and the second one as QTFT.

\section{Infinitely Repeated Quantum PD}
We are now set up to investigate the repeated quantum PD. While more general results and monotonous calculations have been shelved in the appendices, we present the eclectic results in this section using the interaction between the selected aforementioned strategies. The repeated PD game considered is supposed to be played for a large number of (theoretically, infinite) rounds. Note that the discount factor will always be taken to be less than unity. However, as we are interested in the effect of entanglement on the game, we shall allow $\varepsilon$ to take all possible allowed values.

\subsection{Efficiency of CTFT}
\label{subsec: ctft}
Since unlike its classical counterpart, the repeated quantum PD is like a stochastic game, the standard results for TFT-ALLD case (as mentioned in Sec.~\ref{sec:CG}) does not carry over to it. We depict below the circuit diagram where CTFT faces any repeated quantum strategy ALL-${\sf U}$ where ${\sf U}$ can either be defect (${\sf D}$) or Hadamard move ($\sf H$) or supercooperation (${\sf Q}$) or any other valid action:
\vskip 0.25cm
 \hskip 1.0em
\Qcircuit @C=0.7em @R=1.2em {
\lstick{\vert 0 \rangle} & \multigate{1}{ {\sf J} (\varepsilon)} & \gate{{\sf C}} & \gate{{\sf U}} & \qw & {\ldots} & \hskip 1em  & \gate{{\sf U}} & \multigate{1}{ {\sf J}^{\dagger} (\varepsilon)} &  \qw &\rstick{\raisebox{-2.9em}{$\vert \psi_m\rangle.$}}  \\
\lstick{\vert 0 \rangle} & \ghost{ {\sf J} (\varepsilon)} & \gate{{\sf U}} & \gate{{\sf U}} & \qw & {\ldots} & \hskip 1em & \gate{{\sf U}} & \ghost{{\sf J}^{\dagger} (\varepsilon)} & \qw }
\vskip 0.25cm
CTFT's interaction with ALLC and ALLQ is not very interesting: For the former, all the elements of the payoff matrix are identically same and hence there is no strict NE; for the latter the payoff matrix is
  \begin{eqnarray*}  
	\centering
    \begin{tabular}{cc|c|c|}
      & \multicolumn{1}{c}{} & \multicolumn{2}{c}{}\\
      & \multicolumn{1}{c}{} & \multicolumn{1}{c}{CTFT}  & \multicolumn{1}{c}{ALLQ} \\\cline{3-4}
      \multirow{2}*{  {${\sf A}$}=}  & CTFT & $\frac{R}{1-w}$ & $\frac{P \sin^2 2\varepsilon+R \cos^2 2\varepsilon +wR}{1-w^2}$ \\\cline{3-4}
      & ALLQ & $\frac{P\sin^2 2\varepsilon + R \cos^2 2\varepsilon +wR}{1-w^2}$ & $\frac{R}{1-w}$ \\\cline{3-4}
    \end{tabular}
    \label{table:CTFT-ALLQ}
\end{eqnarray*}
 and hence, evidently, both CTFT and ALLQ are always NE---irrespective of the values of $\varepsilon$ and $w$.

 With ALLD in picture, the payoff matrix becomes
  \begin{eqnarray*}  
	\centering
    \begin{tabular}{cc|c|c|}
      & \multicolumn{1}{c}{} & \multicolumn{2}{c}{}\\
      & \multicolumn{1}{c}{} & \multicolumn{1}{c}{CTFT}  & \multicolumn{1}{c}{ALLD} \\\cline{3-4}
      \multirow{2}*{  {${\sf A}$}=}   & CTFT & $R/(1-w)$ & $(S+wT)/(1-w^2)$ \\\cline{3-4}
      & ALLD & $(T+wS)/(1-w^2)$ & $ (P+wR)/(1-w^2)$ \\\cline{3-4}
    \end{tabular}
    \label{table:CTFT-ALLD}
\end{eqnarray*}
 which immediately implies that the conditions for CTFT and ALLD to be strict NE respectively are
\begin{subequations}
\begin{eqnarray}
&&w >w^{CTFT}_{ALLD}\equiv \frac{T-R}{R-S},\label{eq:cd}\\
{\rm and}\,\,&& w < w^{ALLD}_{CTFT}\equiv\frac{P-S}{T-R}.
\label{eq:dc}
\end{eqnarray}
\end{subequations}
An intriguing observation is that quite opposite to what happens in the classical case, here the sucker's payoff ($S$) and not punishment ($P$) has a role in deciding whether CTFT is NE or not. One can note from inequality (\ref{eq:cd}) that CTFT is always NE above a threshold discount factor, $w^{CTFT}_{ALLD}$, since because of the condition $2R>T+S$, $(T-R)/(R-S)\in(0,1)$. However, ALLD is a NE irrespective of the value of the discount factor for all those games for which $R+P\ge S+T$ because $(P-S)/(T-R)\ge1$ in this case [refer inequality (\ref{eq:dc})]; whereas, for other games, one can always find some discount factor for which  ALLD is NE because $w^{ALLD}_{CTFT}\in(0,1)$. The necessary condition for the coexistence of both strategies as NE is $w^{ALLD}_{CTFT}>w^{CTFT}_{ALLD}$, i.e.,  $(T-R)^2<(R-S)(P-S)$. Clearly, this coexistence ceases when temptation, $T>R+\sqrt{(R-S)(P-S)}$.

\begin{figure}
   \begin{center}
   \includegraphics[width=0.48\textwidth]{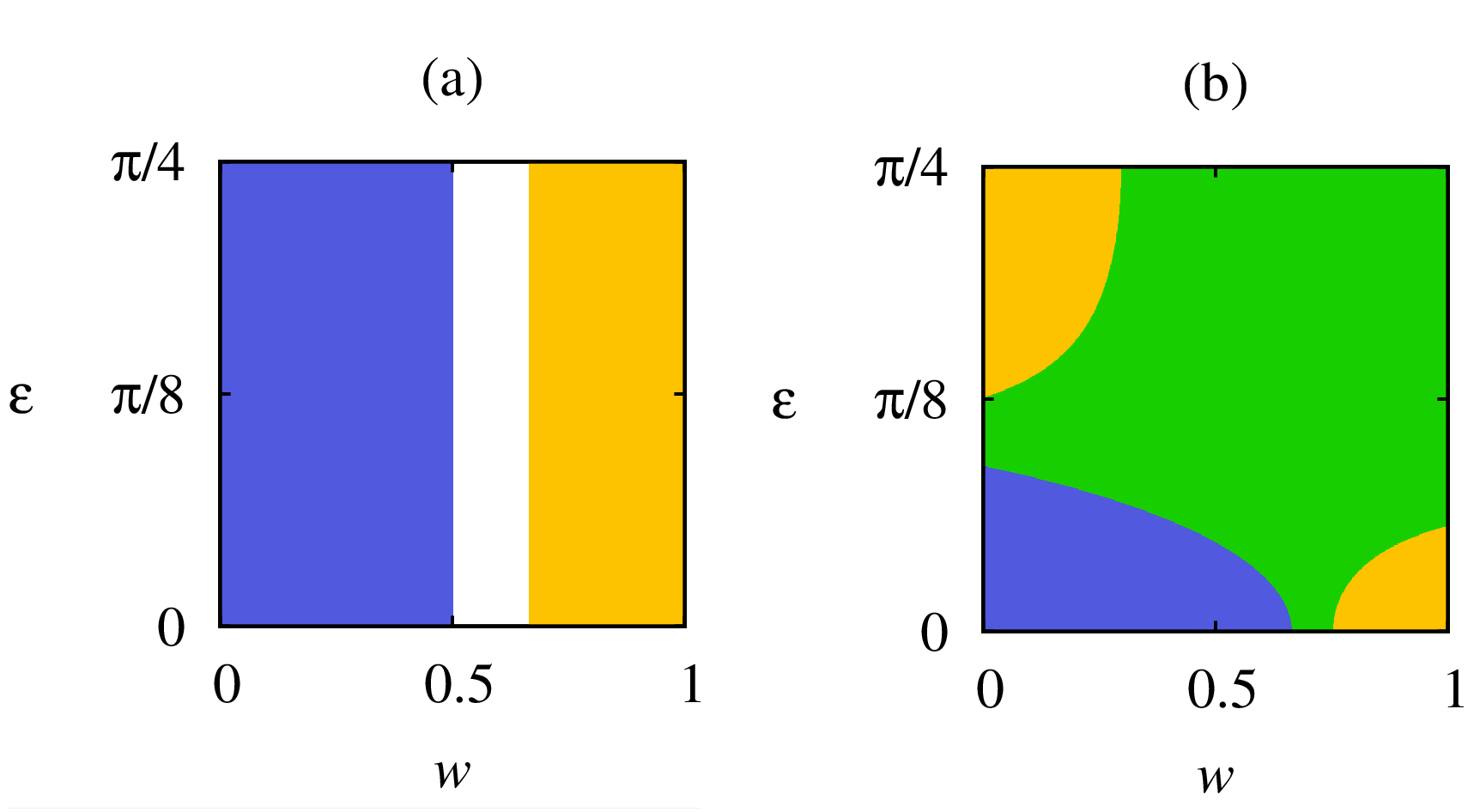}
    \end{center}
   \caption{{We depict the case CTFT-ALLD (in subplot (a)) and CTFT-ALLH (in subplot (b)) for the case when $R=3, S=0, T=5, P=1$. The region where CTFT is strict NE is marked in yellow and the regions where ALLD and ALLH are strict NE are marked in blue  in (a) and (b) respectively.The regions marked with green and white depict where both the strategies are strict NE and none of them are strict NE respectively.}}%
  \label{fig:ctft} 
   \end{figure}

We illustrate our findings in Fig.~\ref{fig:ctft}. A quick comparison between  Fig.~\ref{fig_classical} and  Fig.~\ref{fig:ctft}a reveals the following interesting observations: In quantum repeated PD, (a) ALLD is less advantageous because now it is strict NE only for the discount factor below a threshold value, (b) unlike its classical counterpart, ALLD and CTFT may not coexist as strict NE for some PD games, (c) there may be a range of discount factor where neither CTFT nor ALLD is strict NE, and (d) CTFT can be the only strict NE above a threshold discount factor.

The scenario where a CTFT player encounters an opponent playing the Hadamard move $\sf H$ repeatedly has not classical counterpart. The payoff matrix accounting for such interaction takes the following form:
\begin{widetext}
  \begin{eqnarray*}  
	\centering
    \begin{tabular}{cc|c|c|}
      & \multicolumn{1}{c}{} & \multicolumn{2}{c}{}\\
      & \multicolumn{1}{c}{} & \multicolumn{1}{c}{CTFT}  & \multicolumn{1}{c}{ALLH} \\\cline{3-4}
      \multirow{2}*{  {${\sf A}$}=}    & CTFT & $ \frac{R}{1-w}$ & $\frac{\cos^2 2\varepsilon \left[R(1+w) + S+wT \right] + \sin^2 2\varepsilon\left[P(1+w) + wS+ T \right]}{2(1-w^2)}$ \\\cline{3-4}
      & ALLH & $\frac{\cos^2 2\varepsilon \left[R(1+w) + T+wS \right] + \sin^2 2\varepsilon \left[P(1+w) + wT+S \right]}{2(1-w^2)}$ & $\frac{(R+S+T+P)+4wR}{4(1-w^2)}$ \\\cline{3-4}
    \end{tabular}
    \label{table:CTFT-ALLM}
\end{eqnarray*}
\end{widetext} 
which suggests that the conditions for which CTFT and ALLH are strict NE are 
\begin{subequations}
\begin{eqnarray}\label{eq:cam}
&& w \left(2R - (R+S) \cos^2 2\varepsilon - (P+T) \sin^2 2\varepsilon \right) \nonumber\\
&&>(R+T) \cos^2 2\varepsilon + (P+S) \sin^2 2\varepsilon- 2R 
\end{eqnarray}
and
\begin{eqnarray}
&&w\left(2R - (R+T) \cos^2 2\varepsilon -(P+S)\sin^2 2\varepsilon\right)\nonumber\\ &&> (R+S) \cos^2 2\varepsilon + (P+T) \sin^2 2\varepsilon - \frac{R+S+T+P}{2} \nonumber\\\label{eq:6b}
\end{eqnarray}
\end{subequations}
respectively. An interesting point may be noted in the limit of zero entanglement: On comparing inequalities (\ref{eq:cam}) with (\ref{eq:cd}), we find them to be identical at $\varepsilon=0$. In other words, the range of discount factor for which CTFT is strict NE turns out to be exactly the same as in the case of CTFT vs. ALLD. Moreover, in this zero entanglement limit, in the particular games where $R+P \ge S+T$, ALLH is NE whatever be the value of discount factor. This is because in inequality~Eq.(\ref{eq:6b}) when $\varepsilon=0$ is put, for the aforementioned class of games, the inequality is always satisfied for any value of $w\in(0,1)$.

The complex dependence of the region corresponding to strict NE on both the entanglement and the discount factor may be showcased through a specific game considered in caption.~\ref{fig:ctft}b.  It is evident from the plot that for the high enough entanglement, CTFT is strict NE irrespective of the discount factor whereas the same is true for ALLH only in an intermediate band of entanglement values. However, for the low values of entanglement, CTFT (and ALLH) are strict NE for a discount factor above (and below) a threshold value. The dependence on the discount factor is similar:  For the high enough discount factor, CTFT is strict NE irrespective of the entanglement whereas the same is true for ALLH only in an intermediate band of discount factor. There is a fairly large region of coexisting NE as well. In summary, Hadamard move which can beat classical strategies can be beaten in quantum repeated games.
   \subsection{Efficiency of  QTFT}
We exhibit below the circuit diagram where QTFT confronts any repeated quantum strategy ALL-${\sf U}$ where ${\sf U}$ can either be defect (${\sf D}$) or Hadamard move (${\sf H}$) or supercooperation (${\sf Q}$) or any other allowed action:
\vskip 0.25cm
 \hskip 1.0em
\Qcircuit @C=0.7em @R=1.2em {
\lstick{\vert 0 \rangle} & \multigate{1}{ {\sf J} (\varepsilon)} & \gate{{\sf Q}} & \gate{{\sf U}} & \qw & {\ldots} & \hskip 1em  & \gate{{\sf U}} & \multigate{1}{ {\sf J}^{\dagger} (\varepsilon)} &  \qw &\rstick{\raisebox{-2.9em}{$\vert \psi_m\rangle.$}}  \\
\lstick{\vert 0 \rangle} & \ghost{ {\sf J} (\varepsilon)} & \gate{{\sf U}} & \gate{{\sf U}} & \qw & {\ldots} & \hskip 1em & \gate{{\sf U}} & \ghost{{\sf J}^{\dagger} (\varepsilon)} & \qw }
\vskip 0.25cm
QTFT's interaction with ALLQ is trivial as all the elements of the corresponding payoff matrix are identically same and hence there is no strict NE.   With ALLC, the payoff matrix turns out to be
  \begin{eqnarray*}  
	\centering
    \begin{tabular}{cc|c|c|}
      & \multicolumn{1}{c}{} & \multicolumn{2}{c}{}\\
      & \multicolumn{1}{c}{} & \multicolumn{1}{c}{QTFT}  & \multicolumn{1}{c}{ALLC} \\\cline{3-4}
      \multirow{2}*{  {${\sf A}$}=}    & QTFT & $\frac{R}{1-w}$ & $\frac{R \cos^2 2\varepsilon+P \sin^2 2\varepsilon}{1-w}$ \\\cline{3-4}
      & ALLC & $\frac{R \cos^2 2\varepsilon+P \sin^2 2\varepsilon}{1-w}$ & $\frac{R}{1-w}$ \\\cline{3-4}
    \end{tabular}
    \label{table:QTFT-ALLC}
  \end{eqnarray*}  
  which implies that irrespective of the values of $\varepsilon$ and $w$, both QTFT and ALLC are always NEs.
  
  In view of the above, It may be impulsive to guess that the fate of QTFT would be same as---or at least analogous to---CTFT given that mutual supercooperation and mutual cooperation fetch the same payoff; however, this is not the case. Remember that QTFT has no classical counterpart. Let us now observe how it fares against ALLD and ALLH.
  
We consider the following payoff as obtained after relevant calculations (vide Appendices): 
 \begin{widetext}
  \begin{eqnarray*}  
	\centering
    \begin{tabular}{cc|c|c|}
      & \multicolumn{1}{c}{} & \multicolumn{2}{c}{}\\
      & \multicolumn{1}{c}{} & \multicolumn{1}{c}{QTFT}  & \multicolumn{1}{c}{ALLD} \\\cline{3-4}
      \multirow{2}*{  {${\sf A}$}=}    & QTFT & $\frac{R}{1-w}$ & $\frac{(S+wT) \cos^2 2\varepsilon + (wS+T) \sin^2 2\varepsilon}{1-w^2}$ \\\cline{3-4}
      & ALLD & $\frac{(T+wS) \cos^2 2\varepsilon + (wT+S) \sin^2 2\varepsilon}{1-w^2}$ & $\frac{P+wR}{1-w^2}$ \\\cline{3-4}
    \end{tabular}
    \label{table:QTFT-ALLD}
  \end{eqnarray*}  
  \end{widetext}
  and infer that the conditions for QTFT and ALLD to be strict NE are respectively given as
\begin{subequations}
\begin{eqnarray}
&& w\left(R - S \cos^2 2\varepsilon - T \sin^2 2\varepsilon\right) > T \cos^2 2\varepsilon + S \sin^2 2\varepsilon -R, \nonumber \\
\label{eq: qd}
 \end{eqnarray}
 and
 \begin{eqnarray}
&& w\left(R -S \sin^2 2 \varepsilon  -T \cos^2 2 \varepsilon\right) > S \cos^2 2 \varepsilon + T \sin^2 2 \varepsilon - P. \nonumber \\  
\label{eq: dq}
\end{eqnarray}
\end{subequations}

        \begin{figure}
   \begin{center}
   \includegraphics[width=0.48\textwidth]{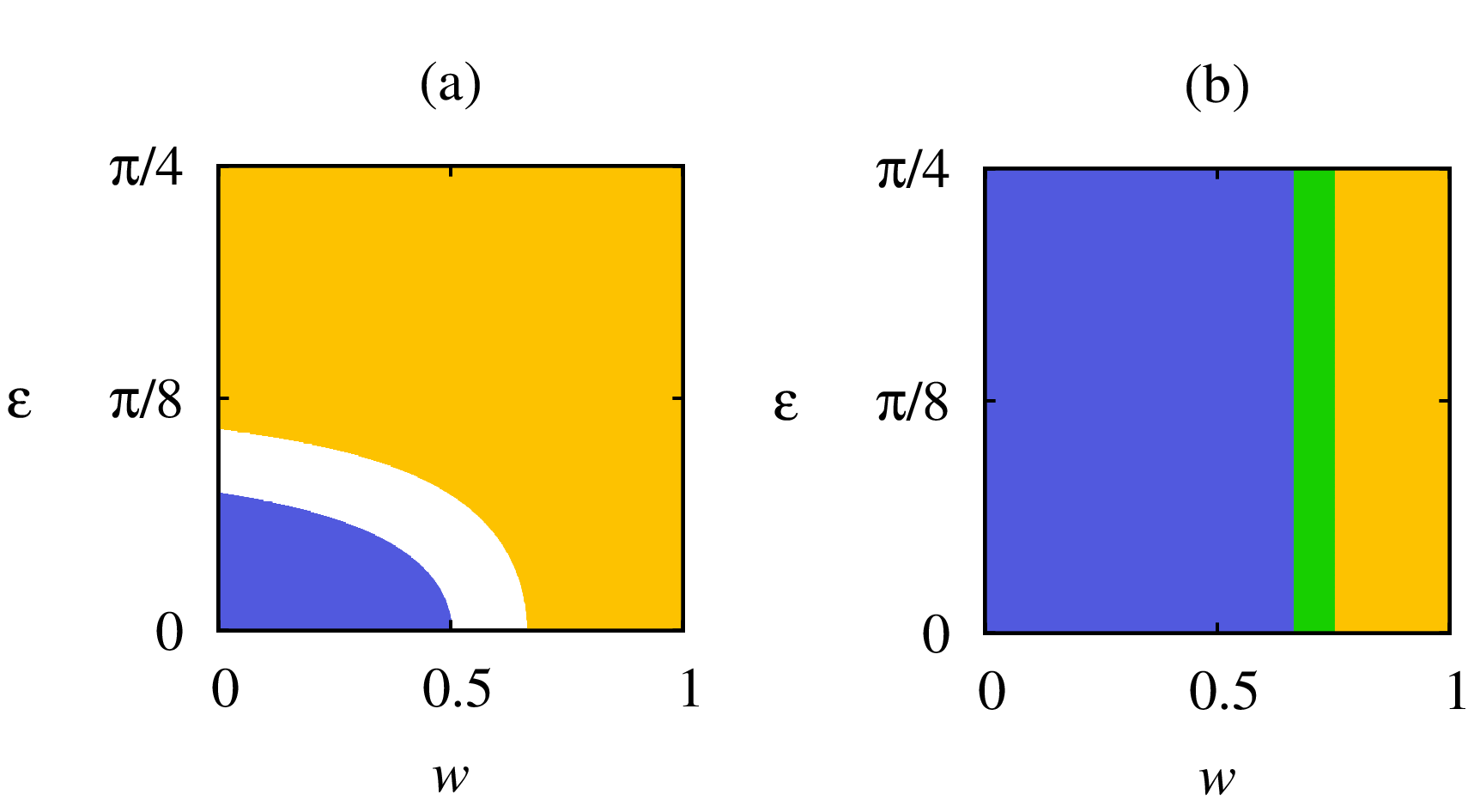}
    \end{center}
   \caption{{We depict the case QTFT-ALLD (in subplot (a)) and QTFT-ALLH (in subplot (b)) for the case when $R=3, S=0, T=5, P=1$. The region where QTFT is strict NE is marked in yellow and the regions where ALLD and ALLH are strict NE are marked in blue in (a) and (b) respectively. The regions marked with green and white depict where both the strategies are strict NE and none of them are strict NE respectively.}}%
  \label{fig:qtft} 
   \end{figure} 
It is worth noting that inequalities (\ref{eq: qd}) and (\ref{eq: dq}) turn out to be identical to inequalities (\ref{eq:cd}) and (\ref{eq:dc}), respectively, in the limit of zero entanglement. For the maximum entanglement, QTFT is always NE---irrespective of the value of discount factor---because $2R>S+T$ holds for repeated PD; additionally, the games for which ALLD is never a NE satisfies $R+P \le S+T$ because inequality~(\ref{eq: dq}) is not satisfied with any value of $w$.

We present this result graphically in Fig.~(\ref{fig:qtft})a. Here, unlike when ALLD played against CTFT, we get a coupled dependence on both the discount factor and entanglement for being NE. For the high enough entanglement, QTFT is strict NE irrespective of the discount factor whereas ALLD can't be NE. However, for the low values of entanglement, QTFT (and ALLH) are strict NE for a discount factor above (and below) a threshold value of the discount factor. In summary, we can safely say that by and large QTFT beats ALLD as an efficient rational strategy.

Against ALLH, the behaviour is again unlike that of CTFT. It is evident from the following payoff matrix
  \begin{eqnarray*}  
	\centering
    \begin{tabular}{cc|c|c|}
      & \multicolumn{1}{c}{} & \multicolumn{2}{c}{}\\
      & \multicolumn{1}{c}{} & \multicolumn{1}{c}{QTFT}  & \multicolumn{1}{c}{ALLH} \\\cline{3-4}
         \multirow{2}*{  {${\sf A}$}=}   & QTFT & $\frac{R}{1-w}$ & $\frac{R(1+w)+S+wT}{2(1-w^2)}$ \\\cline{3-4}
      & ALLH & $\frac{R(1+w)+T+wS}{2(1-w^2)}$ & $\frac{\frac{1}{4}(R+S+T+P)+wR}{1-w^2}$ \\\cline{3-4}
    \end{tabular}
    \label{table:QTFT-ALLM}
  \end{eqnarray*}  
which yields the following conditions for QTFT and ALLH to be strict NE
\begin{subequations}
\begin{eqnarray}
w > \frac{T-R}{R - S}\label{eq:qaa}
\end{eqnarray}
and
\begin{eqnarray}
w < \frac{T+P-R-S}{2(T - R)}   \label{eq:qab}
\end{eqnarray}
\end{subequations}
respectively. It is obvious that the conditions are independent of the entanglement parameter, unlike the CTFT vs. ALLH case. Inspection of inequalities~(\ref{eq:qaa}) and (\ref{eq:cd}) reveals that for whichever set of parameters, CTFT is NE when confronted with ALLH, QTFT would also be so against ALLH. Again comparision between inequalities~(\ref{eq:qab}) and (\ref{eq:6b} confirms that in the absence of entanglement, the condition for ALLH to be NE is identical in the respective cases. This is compatible with the fact that when $\varepsilon=0$, the rows and the columns corresponding to actions $\sf C$ and $\sf Q$ of the two-player four-action payoff matrix (see Sec.~\ref{sec: strategies}) are identical.

Furthermore, we showcase the results for the same particular game in Fig.~\ref{fig:qtft}b. For details, in this game, ALLH is a strict NE only for a discount factor below a threshold whereas the exact opposite is true for QTFT. There is a small range of $w$ for which both of them are simultaneously strict NE. 

\subsection{Classical vs. quantum strategies}\label{sec:cvq}
After having found the fate of TFT strategies, we now turn to compare how repeatedly playing an exclusively quantum strategy ($\sf Q$ and $\sf H$) competes against a repeatedly played classical strategy. We know that in one-shot scenario both the quantum strategies fetches more payoff than the classical ones. However, due to the inherent stochastic game nature of the repeated quantum game, the condition for the quantum strategies to be NE should have non-trivial dependence on the discount factor and the entanglement parameter.

While the case of ALLQ vs. ALLC is rather simple as seen from the payoff matrix below
  \begin{eqnarray*}  
	\centering
    \begin{tabular}{cc|c|c|}
      & \multicolumn{1}{c}{} & \multicolumn{2}{c}{}\\
      & \multicolumn{1}{c}{} & \multicolumn{1}{c}{ALLC}  & \multicolumn{1}{c}{ALLQ} \\\cline{3-4}
               \multirow{2}*{  {${\sf A}$}=}   & ALLC & $\frac{R}{1-w}$ & $\frac{P \sin^2 2\varepsilon +R \cos^2 2\varepsilon + Rw}{1-w^2}$ \\\cline{3-4}
      & ALLQ & $\frac{P\sin^2 2\varepsilon + R \cos^2 2\varepsilon + wR}{1-w^2}$ & $\frac{R}{1-w}$ \\\cline{3-4}
    \end{tabular}
    \label{table:ALLC-ALLQ}
  \end{eqnarray*}  
  which shows that both of them are always NE, the case of ALLQ vs. ALLD is far more interesting: Consider the payoff matrix for the latter case:
  \begin{eqnarray*}  
	\centering
    \begin{tabular}{cc|c|c|}
      & \multicolumn{1}{c}{} & \multicolumn{2}{c}{}\\
      & \multicolumn{1}{c}{} & \multicolumn{1}{c}{ALLD}  & \multicolumn{1}{c}{ALLQ} \\\cline{3-4}
               \multirow{2}*{  {${\sf A}$}=}  & ALLD & $\frac{P + wR}{1-w^2}$ & $\frac{S \sin^2 2\varepsilon +T \cos^2 2\varepsilon + wR}{1-w^2}$ \\\cline{3-4}
      & ALLQ & $\frac{T \sin^2 2\varepsilon +S \cos^2 2\varepsilon + wR}{1-w^2}$ & $\frac{R}{1-w}$ \\\cline{3-4}
    \end{tabular}
    \label{table:ALLD-ALLQ}
  \end{eqnarray*}

\begin{figure}
   \begin{center}
   \includegraphics[width=0.26\textwidth]{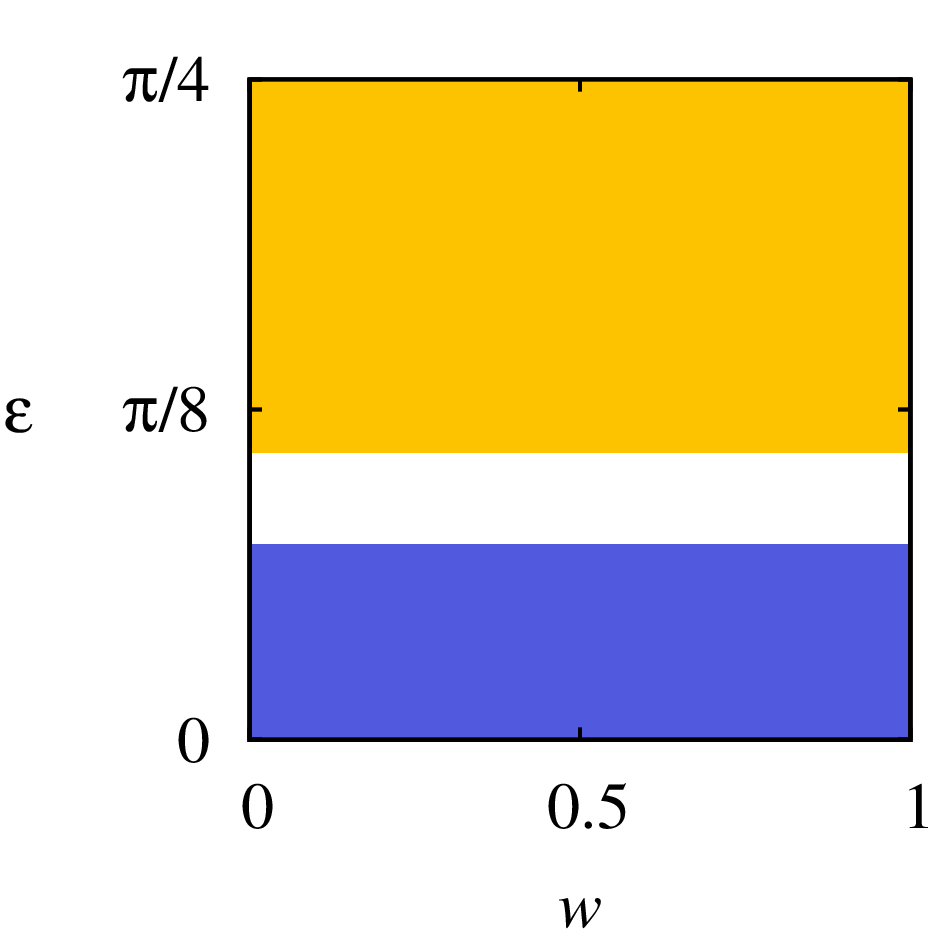}
    \end{center}
   \caption{{We depict the case ALLQ-ALLD for the case when $R=3, S=0, T=5, P=1$. The regions where ALLQ and ALLD are strict NE are marked in yellow and blue respectively. The white region marks where none of the strategies is strict NE.}}%
  \label{fig:allq} 
   \end{figure}

  \begin{table*}
    \setlength{\extrarowheight}{2pt}
    \begin{tabular}{cc|c|c|}
      & \multicolumn{1}{c}{} & \multicolumn{2}{c}{}\\
      & \multicolumn{1}{c}{} & \multicolumn{1}{c}{ALLC}  & \multicolumn{1}{c}{ALLH} \\\cline{3-4}
      \multirow{2}*{  {${\sf A}$}=}  & ALLC & $\frac{R}{1-w}$ & $\frac{(R+S) \cos^2 2\varepsilon + (T+P) \sin^2 2\varepsilon+2wR}{2(1-w^2)}$ \\\cline{3-4}
      & ALLH & $\frac{(R+T) \cos^2 2\varepsilon + (S+P) \sin^2 2\varepsilon+2wR}{2(1-w^2)}$ & $\frac{\frac{1}{4}(R+S+T+P)+wR}{1-w^2}$ \\\cline{3-4}
    \end{tabular}
    \caption{Schematic payoff-matrix of a repeated quantum PD game with ALLC and ALLH strategy.}
    \label{table:ALLC-ALLM}
  \end{table*} 
  
which specifies the conditions for which ALLD and ALLQ are strict NE to be respectively
\begin{subequations}
\begin{eqnarray}
\sin^2 2\varepsilon < \frac{P-S}{T-S},
 \end{eqnarray}
 and
 \begin{eqnarray}
\sin^2 2\varepsilon > \frac{T-R}{T-S}.
\end{eqnarray}
\end{subequations}
We showcase this result in Fig.~\ref{fig:allq}. Clearly, here the condition of strict NE depends only on the entanglement parameter. One can see that ALLD (or ALLQ) is always NE below (or above) a certain value of entanglement. Both can be simultaneously NE for games with $R+P>T+S$ because for such games: $(T-R)/(T-S)<(P-S)/(T-S)$. In passing, we remark that this coexistence region is absent in Fig.~\ref{fig:allq} owing to the particular choice of payoff matrix: Had we chosen a different payoff matrix satisfying the condition, $R+P>T+S$, one would have seen a green region in the plot.

Similar consideration with ALLH get us following conditions \begin{subequations}
\begin{eqnarray}
&& \sin^2 2\varepsilon < \frac{P-S}{(T-S)+(R-P)},
\end{eqnarray}
and
\begin{eqnarray}
&& \sin^2 2\varepsilon > \frac{1}{2}.\label{eq:1/2a}
\end{eqnarray}    
\end{subequations}
for which ALLD and ALLH are respectively strict NE. There always exists a continuous interval of entanglement values, bounded below by zero, such that ALLD is always NE for any value of discount factor. We also observe that one can find a range of entanglement values in which ALLD and ALLH can coexist as NE's for games satisfying the inequality: $3P>T+R+S$. What surprises us more is the fact that ALLH is a NE irrespective of the underlying game structure---all that has to be ensured is $\varepsilon>\pi/8$ [see inequality~(\ref{eq:1/2a})]. 

Similar considerations yield that ALLC and ALLH are NE respectively when
\begin{subequations}
\begin{eqnarray}
&& \sin^2 2\varepsilon > \frac{T-R}{(T-S)+(R-P)},
\end{eqnarray} 
and
\begin{eqnarray}
\sin^2 2\varepsilon < \frac{1}{2}.\label{eq:1/2}
\end{eqnarray}    
\end{subequations}
for which ALLC and ALLH are respectively NE. Clearly, a continuous interval of entanglement values, bounded above by $\pi/4$, exists such that ALLC is always NE. Furthermore, a range of entanglement values exists in which ALLC and ALLH can coexist as NE's for games satisfying the inequality: $3R>T+P+S$. It is quite interesting that ALLH is a NE irrespective of the underlying game structure whenever $\varepsilon<\pi/8$ [see inequality~(\ref{eq:1/2})]. 
        \begin{figure}
   \begin{center}
   \includegraphics[width=0.48\textwidth]{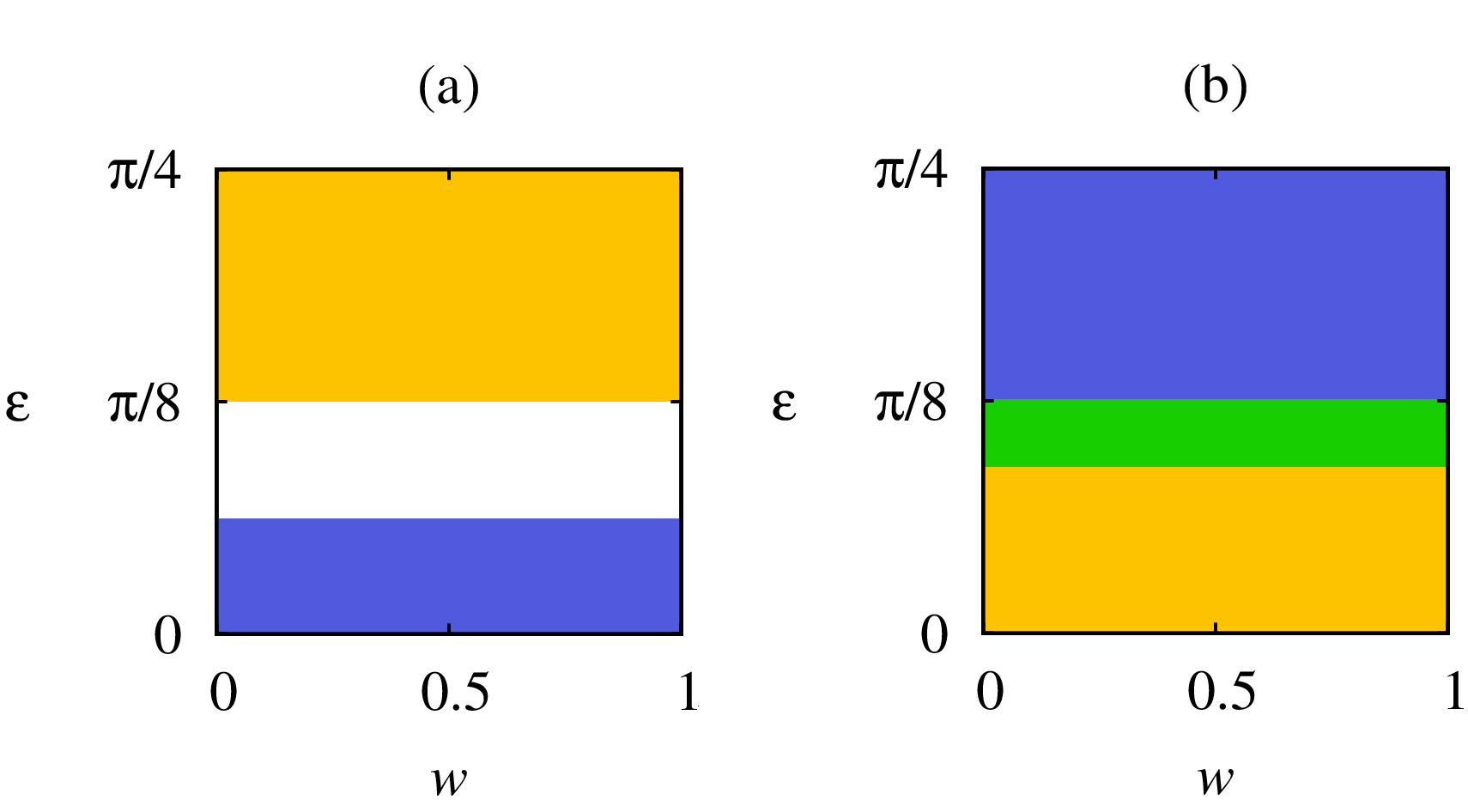}
    \end{center}
   \caption{{We depict the case ALLH-ALLD (in subplot (a)) and ALLH-ALLC (in subplot (b)) for the case when $R=3, S=0, T=5, P=1$. The region where ALLH is strict NE is marked in yellow and the regions where ALLD and ALLC are strict NE are marked in blue in (a) and (b) respectively. The regions marked with green and white depict where both the strategies are strict NE and none of them are strict NE respectively.}}%
  \label{fig:allm} 
   \end{figure}

Thus, the fate of ALLH against the classical strategies is independent of the discount factor. Higher entanglement is favourable for ALLH (as strict NE) whereas, in the lower entanglement, ALLD is advantageous. On the contrary, when competing with ALLC,  a higher value of entanglement makes ALLC NE whereas, for lower values of entanglement favours ALLH.

\section{A Deceptive Triviality}
At first glance, if not much thought is put in, it may appear that if an action (say $A_1$) triumphs another action (say $A_2$) in a one-shot game, then ALL-$A_1$ would beat ALL-$A_2$ as well (as has been witnessed in Sec.~\ref{sec:cvq}) since the corresponding repeated game appears like a trivial sequence of identical one-shot interactions between $A_1$ and $A_2$. This intuition might have been true in the classical games, but in the quantum games this intuition is  fallacious because---by virtue of being effectively a stochastic game---the same action-profile need not lead to the same payoffs at every round of repeated quantum game. Hence what appears like a trivial result actually begs a non-trivial explanation for it. Let us discuss that now.

Quantum games have many interesting distinctions from classical games---an important one being the access to SU($2$) strategies. It is well-known that SU($2$) operations are actually rotations on a Bloch sphere. One important aspect of such rotations is their additive property: When the subsequent rotations commute, (i.e., the only case when all subsequent rotations are of the same magnitude about a fixed axis), the action of $m$ subsequent rotations each by an angle $\theta$ is equivalent to a rotation by an angle $m\theta$. Using this property one can write $|\psi_m\rangle$ (of Eq.~\ref{eq: psi_n}) in the following form:
\begin{eqnarray}
|\psi_m\rangle={\sf J}^\dagger \left[{\sf U}^A(m\theta_A,\alpha_A,\phi_A) \otimes {\sf U}^B(m\theta_B,\alpha_B,\phi_B) \right] {\sf J}|\psi_0\rangle. \nonumber \\
\label{eq: psi_n_mtheta}
\end{eqnarray}
As clarified below, this expression hints towards a possibility of strategy-induced periodicity or aperiodicity in our protocol for the repeated quantum game. 

It is easy to note that for the case where both $\theta^A/(2\pi)$ and $\theta^B/(2\pi)$ are rational---i.e., $\theta^A= (p^A/q^A)2\pi $ and $\theta^B= (p^B/q^B)2\pi $ (where, $p^A$, $q^A$, $p^B$, and $q^B$ are integers; and $q^A$ and $q^B$ are not zero)--- the initial quantum environment, $|\psi_0\rangle$, gets repeated after LCM($q^A,q^B$) rounds, i.e., $\vert \psi_{\text{LCM}(q^A,q^B)} \rangle = \vert \psi_0 \rangle$. (Here `LCM' stands for the least common multiple.) For a game between any two repeated actions from the set $\{{\sf C}$, ${\sf D}$, ${\sf Q}$, ${\sf H}\}$, the quantum environment gets repeated after every two rounds. After the first round, the payoffs are the known payoffs of one-shot game whereas in the second round, each action gets the same payoff $wR$ (in our investigations: $\vert \psi_0 \rangle=|00\rangle$); consequently, the NE remains the same for both one-shot and multiple-shot games in such cases. However, we must remark that for the scenario where the repetition of the quantum environment happens after three or more rounds, we can not make any general comment about the multiple-round games from the knowledge of a one-shot game. One case study regarding a repeated game, where the initial state repeats itself after three rounds has been shown in Appendix~\ref{appendix:C}.

Most intriguingly, it is now evident from aforementioned discussion that in the case where at least one of the repeated actions is such that the corresponding $\theta$ can not be written as a rational multiple of $2\pi$, no quantum environment ever gets repeated. This is equivalent to saying that the sequence of payoffs (of each round) in the equivalent stochastic game setup shows aperiodicity. In this scenario, although the cumulative payoff converges, a closed-form expression for it is very difficult---if at all it is possible---to derive. 

\section{Discussion and Conclusions}
Quantum game theory is futuristic: It offers many interesting prospects of applications in a wide range of topics like quantum communication~\cite{nayak2003pra,cleve2010,solmeyer2018jpha,hidalgo2008}, quantum cryptography protocols~\cite{bennett_1984,fitzi2001prl}, quantum gambling~\cite{goldenberg1999prl,hwang2001pra,hwang2002pra,zhang2017npj} etc.  Recent achievements~\cite{sisodia2017pla,kairon2020,solmeyer2018qst, khan2018qip} in the efficient preparation, manipulation, and measurements of quantum states might sooner or later lead to the realization of realistic quantum computers where spatially separated players can select their strategy independently and play games using quantum resources. The realization put forward in this paper that the repeated quantum games can actually be interpreted as a stochastic game will help us to constructively tap into the tools and techniques of the stochastic games to the quantum domain, and to further the prospects of the quantum games in future applications.

Furthermore, it should be emphasized that the classical repeated games are found mostly in the context of evolutionary game theory where it is the concept of evolutionarily stable strategy (ESS)---and not the NE---which plays the main role as a solution concept. Mathematically, a strict NE is ESS~\cite{Weibull_2004}---a reason why we focused solely on strict NE throughout this paper. We have consciously refrained from mentioning ESS explicitly, because of how much quantum game theory is related to evolutionary game theoretic scenarios---usually encountered in biological and social sciences---is not presently very clear to us. Nevertheless, although debatable, some authors argue that quantum game theory may even be useful to socio-economics, biology, and cognitive sciences: e.g., quantum survival games at the molecular level~\cite{arndt2009hfsp,dawkins1976book}, dynamics of evolution~\cite{bomze1996cegore}, and trading in market games~\cite{pitrowski2003,pitrowski2002physica,pitrowski2003physica}, and isomorphism between quantum mechanics and mathematical economics~\cite{lambertini2002}. 

Two remarks on our setup must be made: Firstly, the quantum game theory essentially allows for a continuum set of strategies in a quantum game---by virtue of the possibility of superposition of basic quantum states---even if its corresponding classical counterpart is restricted to have only a countably finite strategy set. Fortunately, in a strategic game with a compact continuum set of strategies, the existence of at least a single Nash equilibrium (NE)~\cite{nash1950pnas, Glicksberg_1952} is guaranteed. Evidently, quantum games are much richer. However, we have only considered pure actions and (repeated game) strategies that are composed of only such actions. Exploration of mixed strategies has remained beyond the scope of our work. Secondly, we emphasize here that the quantum games we have studied are within the paradigm of the Eisert--Wilkens--Lewenstein (EWL) protocol~\cite{eisert1999prl, eisert1999jmodopt}. Among many existing protocols~\cite{marinatto2000pla,piotr2013jpha}, we focus on the EWL protocol because of its wide-spread acceptability~\cite{flitney2006pla,benjamin2001pra,enk2002pra,du2002cpl,luna2017sr,khan2018qip,phoenix2020pla} and simplicity, and for the sake of a concrete testbed of our idea. One would have obtained qualitatively similar results for any other protocol owing to the change of the initial state after each round; therefore correspondence between the strategies and the payoffs also changes. 

The entanglement parameter and the discount factor, mathematically representing the shadow of the future, are intertwined in our theory of repeated quantum games. For a given payoff matrix, they together decide which quantum repeated strategy should be the rational outcome. The results obtained are mostly and understandably different than what is expected in the classical setting. Since the quantum entanglement parameter is externally tuneable and is unaffected by the ongoing game, one may envisage its usage in imposing a desired outcome, i.e., it can help in mechanism design~\cite{Hurwicz_Reiter_2008} in quantum repeated games. We take this opportunity to point out an interesting finding: The difference of incentives arising from deviating to defect, i.e., $(T-R)-(P-S)$, dictates the NE in many of the conflicts discussed in the paper (see Table~\ref{tab:similar_results_2} for summary): $\Delta^{CC}_{DC}\equiv T-R$ is the incentive to deviate to defect in the action profile  (Cooperate, Cooperate), while $\Delta^{CD}_{DD}\equiv P-S$ is the incentive to deviate to defect in the action profile  (Cooperate, Defect).
\begin{table}[h!]
	\centering
	\begin{tabular}{|p{2.7cm}||p{4.9cm}| }
		\hline
		Conflict & if  $\Delta^{CC}_{DC}<\Delta^{CD}_{DD}$ \\
		\hline
		CTFT vs. ALLD & ALLD is NE $\forall w$ and $\forall \varepsilon$ \\
		CTFT vs. ALLH  & ALLH is NE $\forall w; \varepsilon=0$ \\
		QTFT vs. ALLD & ALLD is not NE for any $w$ and any $\varepsilon$ \\
		ALLD vs. ALLQ & coexistence of NE for any $w$ and for a game-dependent $\varepsilon \in [\varepsilon_1, \varepsilon_2]$ where both $\varepsilon_1, \varepsilon_2 \in (0,1)$ \\
		QTFT vs. ALLH & ALLH is NE $\forall w$ and $\forall \varepsilon$ \\
		\hline
	\end{tabular}
	\caption{Dependence of NE on the incentives to deviate.}
	\label{tab:similar_results_2}
\end{table}

Needless to say, our work on quantum repeated games and its interpretation as a stochastic game opens up an avenue for many more explorations in the field of quantum game theory: We have worked only on repeated simultaneous two-player games with reactive strategies. In the future, one can go beyond reactive strategies to memory-k strategies~\cite{chen2017}, simultaneous games to alternating games~\cite{Faigle_2022}, and two-player games to multi-agent games~\cite{Faigle_2022}. However, what intrigues us the most is to investigate if a classical stochastic game may be mapped as a repeated quantum game----converse of what has been established herein.

Before we conclude, we would like to draw the attention of the readers towards another futuristic aspect of quantum games: The interconnection between quantum games, stochastic games, and artificial intelligence. The research community of machine learning and artificial intelligence (AI) would naturally be interested in the realistic reinforcement learning~\cite{kaelbling1996reinforcement} scenario where more than one agent are vying against each other for rewards---a natural stochastic game setup where any focal agent is effectively making decisions in a dynamic environment. Autonomous agents making decisions in unknown dynamic environments has been studied in many artificial intelligence problems, e.g., in chess, robotics, autonomous vehicle control etc.~\cite{Silver2016nature,pieter2006,shwartz2016,ozdaglar_2021}. Interestingly, quantum game theory has been argued to be offering techniques for analysing AI-related topics~\cite{miakisz2006tcs}. Quantum theory considers the objective reality along with the subject and the process of observation. This is in contrast to the theory of classical mechanics. The inconsistency of quantum mechanics with local realism---found in several experiments---suggests that complete objective information about the object is inaccessible to the observer and is prohibited fundamentally by nature~\cite{clauser1974prd,clauser1978rpp}.  Several attempts are being made in classical AI to replicate the human mind using the simulation by digital computation; some have even passed the Turing Test~\cite{oktar2020,noever2022turing}. However, replicating human consciousness remains the final barrier for the AI. Thus, the realization that the subjectivity of mind cannot be explained by classical mechanics, has some to argue---not without their opponents---that quantum phenomena must be invoked~\cite{hameroff2014physlr}. Recent developments in quantum technologies and progresses in designing multi-component quantum computer suggest that quantum protocols would sooner or later take over their classical counterparts where entanglement can be used as a resource~\cite{bennett1992joc,gisin2002rmp,shor1994}. This gives rise to the emergence of a new branch---quantum artificial intelligence---where one asks whether mind activities can be transformed into quantum information processing tasks~\cite{miakisz2006tcs}.

\acknowledgements
  The authors are grateful to Shrestha Biswas for some preliminary calculations and fruitful discussions. SC acknowledges the support from SERB (DST, govt. of India) through project no. MTR/2021/000119. AM acknowledges valuable discussions with Dr. Jens Christian Claussen during his stay as a postdoctoral fellow at the School of Computer Science, UoB where the majority of this work was done. AM is currently employed at NIMHANS, Bengaluru, India.
\appendix

\setcounter{equation}{0}
\renewcommand{\theequation}{A\arabic{equation}}
\section{A Comparative discussion on  three-parameter and two-parameter strategy spaces}

\textcolor{black}{Both the input and the output state of a unitary operation lie on the surface of the Bloch sphere. The form of unitary operation belonging to $SU(2)$ group rotation is of the form as given in Eq.~\ref{eq:3} of the main text (see Sec.~\ref{sec:QG})}. For the most general form~\cite{nielsen2000book} the parameters $\gamma$ and $\delta$-----are further parameterized in terms of three strategy parameters $\theta$,$\alpha$ and $\phi$; where $\gamma = \cos\frac{\theta}{2} - i \cos \alpha \sin\frac{\theta}{2}$ and $\delta = -i e^{-i \phi} \sin \alpha \sin{\frac{\theta}{2}}$. The unitary operator rotates the input state by an angle $\theta$ about the unit vector $(\sin \alpha \cos \phi, \sin \alpha \sin \phi,  \cos \alpha)$ axis of the Bloch sphere. Thus, the quantum strategy space is spanned by the parameters $\theta$, $\alpha$ and $\phi$ where $\theta \in [0,2\pi)$, $\alpha \in [0,\pi)$ and $\phi \in [0,2\pi)$. In summary, we can specify the quantum strategies for Alice and Bob as a pair of triplets ${\bf \Theta}^A=\{\theta_A, \alpha_A, \phi_A\}$ and ${{\bf \Theta}^B}=\{\theta_B, \alpha_B, \phi_B\}$ respectively.
\begin{table}[ht]
\centering
\begin{tabular}{ |p{4cm}||p{1cm}|p{1cm}|  }
 \hline
Strategies & $\theta'$ & $\phi'$ \\
 \hline
{Cooperation {\sf C}}  & $0$ & $0$\\
{Defection {\sf D} or {\sf X}} gate   & -- & --\\
 { {\sf Y}} gate (another possible Defection strategy)&   $\pi$  & $0$\\
{Super-cooperator {\sf Q} }   & $0$ & $\pi/2$ \\
 {Hadamard (Miracle) {\sf M} }  & $\pi/2$ & $\pi/2$\\
  \hline
\end{tabular}
\caption{Parameter values for some standard actions in the conventional two-parameter subspace.}
\label{tab:strategies_2}
\end{table}

\textcolor{black}{However, the construction of a two-parameter subspace of $SU(2)$ from the three-parameter one is not unique; One can therefore find more than one subspaces of $SU(2)$ with constraint(s) put on any of the three parameters. For example, the general three-parameter space can be reduced to the traditional two-parameter strategy space, commonly used in literature~\cite{eisert1999prl}, effectively by setting the constraint $\cos^2 \frac{\theta}{2} + \sin^2 \frac{\theta}{2} (\cos^2 \alpha + \sin^2 \phi \sin^2 \alpha) = 1$ or $\cos^2 \frac{\theta}{2} + \sin^2 \frac{\theta}{2} (\cos^2 \alpha + \cos^2 \phi \sin^2 \alpha) = 1$. This interesting one-to-two mapping is a direct consequence of the abandonment of the global phase factor. The important actions of the game for this two-parameter space are tabulated in Table~\ref{tab:strategies_2}. Kindly refer to Table~\ref{tab:strategies_3} of the main text (see Sec.~\ref{sec: strategies}) for finding their respective forms in the three-parameter space.} One may note that our chosen Defect action, the ${\sf X}$ gate, does not exist in the two-parameter space defined in \cite{eisert1999prl}. This is the sole reason that most of the literature on conventional two-parameter subspace considers the ${\sf Y}$ gate as the Defect action.

\setcounter{equation}{0}
\renewcommand{\theequation}{B\arabic{equation}}
\section{Derivations of payoff matrices}\label{appendix:B}

\subsection{Generic form of the final state and cumulative payoff}
 In this article, the repeated version of the PD game is initialized with $|00\rangle$ basis state (same as in the EWL protocol). The final state obtained after $m$ iterations of one-shot games is given in Eq.~\ref{eq: psi_n} of the main text (see section \ref{sec:QG}). Now we define the following propagators
 \begin{eqnarray}
 \psi^{a' b'}_{ab}(m) = \langle a'b'\vert \prod^m_{k=1} {\sf J}^\dagger (\varepsilon) ({\sf U}^A_k \otimes {\sf U}^B_k) {\sf J} (\varepsilon) \vert ab\rangle,
 \label{eq: final_state}
 \end{eqnarray}
 for the case when the final state after the $m$ round is $|a'b'\rangle$ given the initial state $|a b\rangle$. Thus a simplified notation of the final state after $m$ rounds turns out to be  as follows

 \begin{eqnarray}
  \vert \psi_f\rangle = \sum^1_{a'=0} \sum^1_{b'=0} \psi^{a' b'}_{ab}(m) \vert a' b'\rangle.   
 \end{eqnarray}
 As a result, the cumulative payoffs after $m$ rounds are
 \begin{eqnarray}
     {\rm Payoff}= \sum^m_{k=1}\sum^1_{a'=0} \sum^1_{b'=0} w^{k-1}\vert \psi^{a' b'}_{ab}(k) \vert^2 {\sf A}_{a'b'}.
     \label{eq:gen_propagators}
 \end{eqnarray}
 
 If we consider the special case where both the players are playing the same quantum strategy ${\sf U}={\sf U}^A={\sf U}^B$ the specific form of the propagators turns out to be as follows: 
\begin{widetext}
\begin{subequations}
\begin{eqnarray}
&&\psi^{00}_{00}(m) = C^2_\varepsilon \gamma^2_m + S^2_\varepsilon \gamma^{*2}_m + iC_\varepsilon S_\varepsilon (\delta^2_m - \delta^{*2}_m), \\
&&\psi^{01}_{00}(m) = \psi^{10}_{00}(m) = -C^2_\varepsilon \gamma_m \delta^*_m + S^2_\varepsilon \gamma^{*}_m \delta_m + iC_\varepsilon S_\varepsilon (\gamma^{*}_m \delta_m +  \gamma_m \delta^*_m), \\
&&\psi^{11}_{00}(m) = C^2_\varepsilon \delta^{*2}_m + S^2_\varepsilon \delta^{2}_m + iC_\varepsilon S_\varepsilon (\gamma^{*2}_m - \gamma^{2}_m), \\
&&\psi^{00}_{01}(m) =  C^2_\varepsilon \gamma_m \delta_m - S^2_\varepsilon \gamma^{*}_m \delta^*_m + iC_\varepsilon S_\varepsilon (\gamma_m \delta_m +  \gamma^*_m \delta^*_m), \\
&&\psi^{01}_{01}(m) = |\gamma_m|^2, \\
&&\psi^{10}_{01}(m) = -|\delta_m|^2, \\
&&\psi^{11}_{01}(m) = -C^2_\varepsilon \gamma^{*}_m \delta^*_m + S^2_\varepsilon  \gamma_m \delta_m - iC_\varepsilon S_\varepsilon (\gamma_m \delta_m +  \gamma^*_m \delta^*_m), \\
&&\psi^{00}_{10}(m) =  C^2_\varepsilon \gamma_m \delta_m - S^2_\varepsilon \gamma^{*}_m \delta^*_m + iC_\varepsilon S_\varepsilon (\gamma_m \delta_m +  \gamma^*_m \delta^*_m), \\
&&\psi^{01}_{10}(m) = -|\delta_m|^2, \\
&&\psi^{10}_{10}(m) = |\gamma_m|^2, \\
&&\psi^{11}_{10}(m) = -C^2_\varepsilon \gamma^{*}_m \delta^*_m + S^2_\varepsilon  \gamma_m \delta_m - iC_\varepsilon S_\varepsilon (\gamma_m \delta_m +  \gamma^*_m \delta^*_m), \\
&&\psi^{00}_{11}(m) = C^2_\varepsilon \delta^{2}_m + S^2_\varepsilon \delta^{*2}_m + iC_\varepsilon S_\varepsilon (\gamma^{2}_m - \gamma^{*2}_m), \\
&&\psi^{01}_{11}(m) = \psi^{10}_{11}(m) =  C^2_\varepsilon \gamma^*_m \delta_m - S^2_\varepsilon \gamma_m \delta^*_m -iC_\varepsilon S_\varepsilon (\gamma^{*}_m \delta_m +  \gamma_m \delta^*_m), \\
&&\psi^{11}_{11}(m) =  C^2_\varepsilon \gamma^{*2}_m + S^2_\varepsilon \gamma^{2}_m + iC_\varepsilon S_\varepsilon (\delta^{*2}_m - \delta^{2}_m), 
\end{eqnarray}
 \label{eq: propagators}
\end{subequations}
\end{widetext}
with $C_{\varepsilon} = \cos \varepsilon$, $S_{\varepsilon} = \sin \varepsilon$, $\gamma_m = \cos \frac{m\theta}{2} -i \cos \alpha \sin \frac{m\theta}{2}$,  and $\delta_m = -i e^{-i\phi} \sin \alpha \sin \frac{m\theta}{2}$. One can note that these propagators of Eq.~\ref{eq: propagators} will help us directly compute the payoffs using Eqs.~\ref{eq:gen_propagators} when both the players adopt same strategy throughout the game.

 We depict below a representative study of effectively using Eqs.~\ref{eq: final_state}--\ref{eq: propagators} to calculate the cumulative payoff for a nontrivial case when CTFT meets any repeated quantum strategy ALLU. One can follow the same procedures for all the other interactions.

  \begin{table*}
    \setlength{\extrarowheight}{2pt}
    \begin{tabular}{cc|c|c|}
      & \multicolumn{1}{c}{} & \multicolumn{2}{c}{}\\
      & \multicolumn{1}{c}{} & \multicolumn{1}{c}{ALLC}  & \multicolumn{1}{c}{ALLR\textsubscript{3}} \\\cline{3-4}
      \multirow{2}*{  {${\sf A}$}=}  & ALLC & $\frac{R}{1-w}$ & $\frac{(R+3S)(1+ w) +w^2R}{4(1-w^3)}$ \\\cline{3-4}
      & ALLR\textsubscript{3} & $\frac{(R+3T)( 1+ w) +w^2R}{4(1-w^3)}$  & $\frac{(R+3S+3T+9P)(1+w+w^2)}{16(1-w^3)}$ \\\cline{3-4}
    \end{tabular}
    \caption{Schematic payoff-matrix of a repeated quantum PD game with ALLC and ALLR\textsubscript{3} strategies.}
    \label{table:ALLC-ALLR}
  \end{table*} 

    \begin{table*}
    \setlength{\extrarowheight}{2pt}
    \begin{tabular}{cc|c|c|}
      & \multicolumn{1}{c}{} & \multicolumn{2}{c}{}\\
      & \multicolumn{1}{c}{} & \multicolumn{1}{c}{ALLD}  & \multicolumn{1}{c}{ALLR\textsubscript{3}} \\\cline{3-4}
      \multirow{2}*{  {${\sf A}$}=}  & ALLD & $\frac{P+wR}{1-w^2}$ & $\frac{(T+3P)+ w(R+3S) +w^2R}{4(1-w^3)}$ \\\cline{3-4}
      & ALLR\textsubscript{3} & $\frac{(S+3P)+ w(R+3T) +w^2R}{4(1-w^3)}$  & $\frac{(R+3S+3T+9P)(1+w+w^2)}{16(1-w^3)}$ \\\cline{3-4}
    \end{tabular}
    \caption{Schematic payoff-matrix of a repeated quantum PD game with ALLD and ALLR\textsubscript{3} strategies.}
    \label{table:ALLD-ALLR}
  \end{table*} 

\subsection{Payoff: A case study of CTFT vs ALLU}
For the case of CTFT vs ALLU (kindly refer the quantum circuit diagram of the main text Sec. IIIA),  the final state after $m$ iterations becomes
\begin{eqnarray}
&& \big|\psi_m\big\rangle = {\sf J}^\dagger (\varepsilon)({\sf U}^A \otimes {\sf U}^B)^{m-1} ({\sf C}^A \otimes {\sf U}^B) {\sf J} (\varepsilon) |00\rangle \nonumber\\
&& = {\sf J}^\dagger (\varepsilon)({\sf U}^A \otimes {\sf U}^B)^{m-1} {\sf J}(\varepsilon)  \left[ {\sf J}^\dagger (\varepsilon) ({\sf C}^A \otimes {\sf U}^B) {\sf J} (\varepsilon) |00\rangle \right] \nonumber\\
&& =  {\sf J}^\dagger (\varepsilon)({\sf U}^A \otimes {\sf U}^B)^{m-1} {\sf J}(\varepsilon) \sum_{a,b} g^{ab} \vert ab\rangle
\nonumber\\
\end{eqnarray}
where,  the new parameters are $g^{00} =  C^2_{\varepsilon}\gamma +  S^2_{\varepsilon}\gamma^* $, $g^{01} =  (- \delta^* C^2_{\varepsilon}  +  \delta S^2_{\varepsilon} )$, $g^{10} =  i  C_{\varepsilon} S_{\varepsilon}(\delta + \delta^*)$ and $g^{11} =  i  C_{\varepsilon} S_{\varepsilon}(\gamma^* - \gamma)$.

Now, using the propagators given in Eq.~\ref {eq: propagators},  with both the players playing the same strategy ${\sf U}^A = {\sf U}^B = {\sf U}$, one can write the final state as
\begin{eqnarray}
    \big|\psi_m\big\rangle = \sum_{a',b'} \left(\sum_{a,b} g^{ab} \psi^{a'b'}_{ab}(m-1)\right) \vert a'b'\rangle, 
 \label{ctft_state}   
\end{eqnarray}

For the specific case when $U = D$, one can simplify the above expression using $\gamma = 0$ and $\delta = -1$ (see Table~\ref{tab:strategies_3} in the main text) as
\begin{eqnarray}
\big|\psi_m\big\rangle &&=  \vert 01 \rangle, \mbox{for odd~~} m, \nonumber\\
&& = \vert 10 \rangle, \mbox{for even~~} m.
\end{eqnarray}
Therefore, the cumulative payoff, in this case, is 
\begin{eqnarray}
  \mbox{Payoff} = \frac{{\sf A}_{01} + w{\sf A}_{10}}{1-w^2}. 
\end{eqnarray}

For the other case when ${\sf U} = {\sf H}$, one can simplify the expression in Eq.~\ref{ctft_state}  using $\gamma = -i/\sqrt{2}$ and $\delta = -1/\sqrt{2}$:
\begin{eqnarray}
&&\big|\psi_m\big\rangle= \frac{1}{\sqrt{2}}\left[-\cos{2\varepsilon}\left(i\vert 00 \rangle + \vert 01\rangle\right) -\sin{2\varepsilon}(i\vert 10 \rangle+\vert 11\rangle)\right], \nonumber\\
&& \mbox{for odd~~} m,\nonumber\\
&& \big|\psi_m\big\rangle= \frac{1}{\sqrt{2}}\left(\cos{2\varepsilon}(\vert 00 \rangle - \vert 10\rangle) +\sin{2\varepsilon}(i\vert 01 \rangle+\vert 11\rangle)\right), \nonumber\\
&& \mbox{for even~~} m.
\end{eqnarray}
Therefore, the cumulative payoff in this case is 
\begin{eqnarray}
  &&\mbox{Payoff} = \frac{\cos^2{2\varepsilon}({\sf A}_{00}  + {\sf A}_{01}) + \sin^2{2\varepsilon}({\sf A}_{10}+ {\sf A}_{11})}{2(1-w^2)} \nonumber\\
  && +w\frac{\cos^2{2\varepsilon}({\sf A}_{00}  + {\sf A}_{10}) + \sin^2{2\varepsilon}({\sf A}_{01}+ {\sf A}_{11}))}{2(1-w^2)}. \nonumber\\
\end{eqnarray}

For the cases of CTFT vs CTFT (and also for QTFT vs QTFT), after $m$ rounds, we can trivially find the state to be
\begin{eqnarray}
\big|\psi_m\big\rangle &&=\vert 00 \rangle, \mbox{for all~~} m. \end{eqnarray}
Therefore, the cumulative payoff is ${\sf A}_{00}/(1-w)$. 

Similarly, for the case of ALLD vs ALLD, the state after $m$ rounds would be
\begin{eqnarray}
\big|\psi_m\big\rangle &&=  \vert 11 \rangle, \mbox{for odd~~} m, \nonumber\\
&& = \vert 00 \rangle, \mbox{for even~~} m. 
\end{eqnarray}
Therefore, the cumulative payoff in this case would be $({\sf A}_{11} + w {\sf A}_{00})/(1-w^2)$. 

For the case of ALLH vs ALLH, the state after $m$ rounds would be given as
\begin{eqnarray}
\big|\psi_m\big\rangle &&=  \frac{1}{2}(-\vert00\rangle -i\vert 01 -i\vert 10\rangle + \vert 11 \rangle),~~\mbox{for odd~~} m, \nonumber\\
&& = \vert 00 \rangle,~~\mbox{for even~~} m. 
\end{eqnarray}
Therefore, the cumulative payoff in this case would be $\left[{\sf A}_{00} + {\sf A}_{01} + {\sf A}_{10} + {\sf A}_{11} + 4 w {\sf A}_{00}\right]/4(1-w^2)$. 

\setcounter{equation}{0}
\renewcommand{\theequation}{C\arabic{equation}}
\section{Classical repeated actions against ALLR\textsubscript{3}}\label{appendix:C}
Let us study a repeated game, where the initial quantum environment repeats every third round. Specifically, consider an infinitely repeated game between ALLR\textsubscript{3} (where ${\sf R}_3 \equiv\{\theta=2\pi/3,\alpha=\pi/2\}$) and ALLC or ALLD.

In ALLC vs. ALLR\textsubscript{3} game, which one of these strategies is NE, is decided respectively by the following conditions: 
 \begin{subequations}
\begin{eqnarray}
&& w^2 R - w(T-R) -(T-R)  > 0, \label{eq:cr}
\end{eqnarray}
and
\begin{eqnarray}
&& w^2\left[(T-R)+ (S+3P)\right]+ w\left[(T-R)+3(P-S)\right]\nonumber\\&& > (R-T)+ 3(S-P),\label{eq:rc}
\end{eqnarray}    
\end{subequations}
Clearly, according to inequality~(\ref{eq:cr}) ALLC is NE whenever $w \ge \left(T-R+\sqrt{(T+R)^2 -4R^2}\right)/2$. This implies that for games with payoff elements satisfying $T-3R \ge T^2 -3R^2 +2TR$, ALLC can not be a NE for any value of the discount factor. However, we observe that ALLR\textsubscript{3} is NE irrespective of the value of discount factor since inequality~(\ref{eq:rc}) is trivially satisfied for any value of $w$ in repeated PD game. We remark that for the one-shot version of this game, Cooperation (${\sf C}$) is never a NE. However, for the infinitely repeated version, ALLC can be NE for  discount factor above a certain threshold value.

Similarly, ALLD vs. ALLR\textsubscript{3} gets us the following conditions:
\begin{subequations}
\begin{eqnarray}
&& 3w^3 R + w^2(2R+4P -3T) \nonumber\\&&+ w(P-S-3T-R)+(P-S)  > 0, \label{eq:dr}
\end{eqnarray}
and
\begin{eqnarray}
&& w^2(S+T+3P)+ 3w\left[(T-R)+3(P-S)\right]\nonumber\\&& > (T-R)+ 3(P-S),\label{eq:rd}
\end{eqnarray}    
\end{subequations}
for which ALLD and ALLR\textsubscript{3} are respectively NE. Inequality~(\ref{eq:dr}) hints that for the discount factor above a certain value, it may be possible to get ALLD as a NE---a result that is never possible for the one-shot version of this game. Inequality~(\ref{eq:rd}) can be simplified to show that $w\ge 1/3$ is a sufficient condition for ALLR\textsubscript{3} to be NE. In other cases, it depends on the value of discount factor and the underlying game. However, in the one-shot version of this game, R\textsubscript{3} is NE irrespective of the value of discount factor.

These two examples justify our claim that for interactions among repeated quantum/classical actions, if the quantum environment is repeated after three or more rounds, no conclusion about the NE solution of the infinitely repeated game can be drawn from the results of one-shot version of the game.

\bibliography{Mukhopadhyay_etal_manuscript.bib}
\end{document}